\documentclass{ws-mpla}

\begin{document}


\catchline{}{}{}{}{}

\title{Lowest-order contributions of chiral three-nucleon interactions \\
to pairing properties of nuclear ground states}

\author{T. Duguet}

\address{CEA, Centre de Saclay, IRFU/Service de Physique
Nucl{\'e}aire, F-91191 Gif-sur-Yvette, France \\
National Superconducting Cyclotron Laboratory and Department of
Physics and Astronomy, Michigan State University,
East Lansing, MI 48824, USA \\
thomas.duguet@cea.fr}

\author{T. Lesinski}

\address{Department of Physics and Astronomy, University of Tennessee,
Knoxville, TN 37996, USA \\
Physics Division, Oak Ridge National Laboratory,
Oak Ridge, TN 37831, USA \\
tlesinsk@utk.edu}

\author{K. Hebeler}

\address{TRIUMF, 4004 Wesbrook Mall, Vancouver, BC, V6T 2A3, Canada \\
hebeler@triumf.ca}

\author{A. Schwenk}

\address{ExtreMe Matter Institute EMMI, GSI Helmholtzzentrum f\"ur
Schwerionenforschung GmbH, 64291 Darmstadt, Germany \\
Institut f\"ur Kernphysik, Technische Universit\"at Darmstadt,
64289 Darmstadt, Germany \\
TRIUMF, 4004 Wesbrook Mall, Vancouver, BC, V6T 2A3, Canada \\
schwenk@physik.tu-darmstadt.de}

\maketitle

\begin{abstract}
We perform a systematic study of the odd-even mass staggering
generated using a pairing interaction computed at first order
in low-momentum interactions. Building on previous work including
the (nuclear plus Coulomb) two-nucleon interaction only, we focus
here on the first-order contribution from chiral three-nucleon
forces. We observe a significant repulsive effect from the
three-nucleon interaction. The combined contribution from two- and
three-nucleon interactions accounts for approximately $70 \%$ of the
experimental pairing gaps. This leaves room for higher-order
contributions to the pairing kernel and the normal self-energy that
need to be computed consistently.
\end{abstract}

\section{Introduction}

The nuclear energy density functional (EDF) approach is used to study
medium-mass and heavy nuclei in a systematic
manner~\cite{bender03a}. Currently employed parameterizations of the
EDF provide a satisfactory description of low-energy properties of
known nuclei. However, their empirical character and the spreading of
the results obtained from different parameterizations, as one moves
away from the valley of stability and enters experimentally unexplored
regions, point to the lack of predictive power of today's
calculations.

Our objective is to improve on this situation by developing
non-empirical EDFs constrained explicitly from inter-nucleon
interactions in free space. As a starting point, we performed the
first systematic study of ground states of finite
nuclei~\cite{duguet08a,lesinski09a,duguet09a} using a nuclear EDF
whose pairing part was obtained from low-momentum two-nucleon (NN)
interactions~\cite{bogner03a,bogner07a} at first order. The present
study builds on these results and focuses on the first-order
contributions from chiral three-nucleon (3N) interactions. Details are
reported elsewhere~\cite{lesinski10a}. Higher-order contributions to
the pairing kernel and the normal self-energy are left to future
works.

\begin{figure*}[t]
\begin{center}
\includegraphics[width=0.6\textwidth]{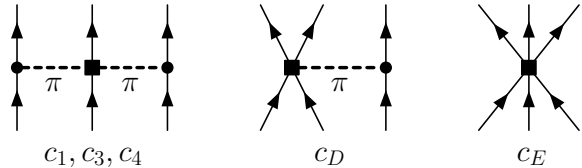}
\end{center}
\caption{Different contributions to chiral 3N forces at N$^2$LO.}
\label{chiralNNN}
\end{figure*}

\section{Method}

Our calculations are based on low-momentum interactions, consisting of
an NN interaction $V_{{\rm low}\,k}$ with a smooth cutoff $\Lambda =
1.8 \, {\rm fm}^{-1}$ and chiral 3N interactions $V_{\text{3N}}$ at
next-to-next-to-leading order (N$^2$LO) with $\Lambda_{\rm 3N} = 2.0
\, {\rm fm}^{-1}$. The different 3N contributions are shown in
Fig.~\ref{chiralNNN}. For details and for the 3N couplings $c_i$,
$c_D$ and $c_E$ used, we refer the reader to Ref.~\cite{bogner09a}.

Starting from the N$^2$LO 3N force, we first construct an
antisymmetrized, density-dependent two-body interaction
$\overline{V}_{\text{NN}}$ by summing one interacting particle over
occupied states in the Fermi sea, extending the calculations of
Ref.~\cite{Hebeler:2009iv} to general isospin
asymmetries. $\overline{V}_{\text{NN}}$ is then added to $V_{{\rm
low}\,k}$ to build the first-order pairing interaction from which
the anomalous self-energy is computed. To proceed, we generate a
high-precision representation of the pairing kernel as a sum of
density-dependent separable terms and employ a local density
approximation. Details will be given in Ref.~\cite{lesinski10a}.

\begin{figure*}[t]
\begin{center}
\includegraphics[width=\textwidth]{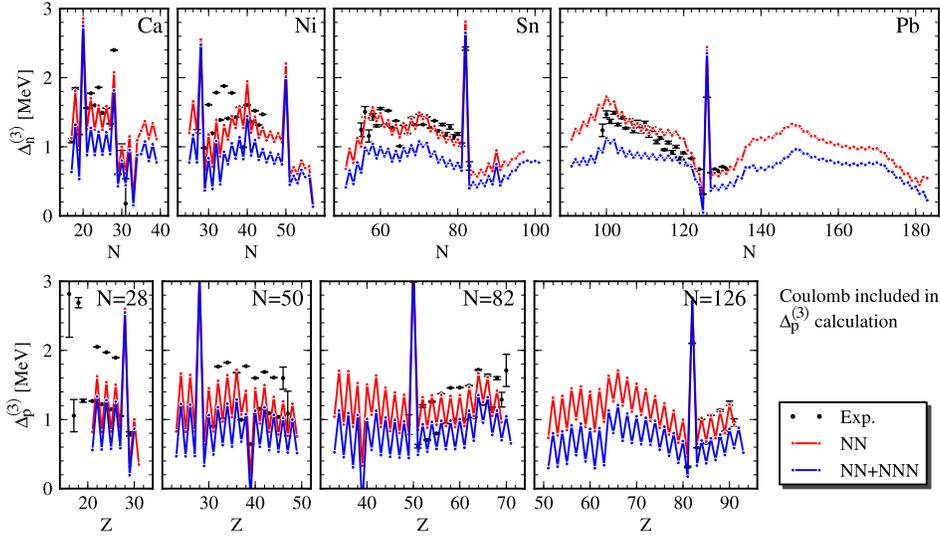}
\end{center}
\caption{Experimental and theoretical neutron/proton three-point mass
differences along semi-magic isotopic/isotonic chains. Results are
obtained from a first-order pairing interaction kernel with and
without 3N contributions.}
\label{fig:gaps:coulomb}
\end{figure*}

The remaining part of the nuclear EDF, i.e., the part accounting for
the normal self-energy that describes the correlated single-particle
motion, is taken as a semi-empirical Skyrme parameterization whose
isoscalar and isovector effective masses have been constrained from
Hartree-Fock calculations of symmetric and pure neutron matter
employing the same NN and 3N low-momentum
interactions~\cite{hebeler09b}. The present results do not depend
significantly on the Skyrme isoscalar effective mass, as long as the
value of the latter at saturation density is $0.67-0.73$.

\section{Results and conclusions}

We limit ourselves to discussing the odd-even mass staggering that
provides a measure of the lack of binding per particle in an odd
isotope/isotone relative to its even neighbors; i.e., it essentially
extracts the pairing gap~\cite{duguet02}. Figure~\ref{fig:gaps:coulomb} displays
theoretical and experimental neutron/proton three-point mass
differences along several semi-magic isotopic/isotonic chains from
proton to neutron drip lines. Results obtained with and without 3N
contributions to the first-order pairing interaction are compared.

The main result obtained with NN only~\cite{duguet09a} is that
theoretical neutron and proton pairing gaps computed at lowest order
are close to experimental ones for a large set of semi-magic spherical
nuclei, although experimental gaps are underestimated in the lightest
systems. The addition of the first-order 3N contribution lowers
pairing gaps systematically by about $30\%$. This is in line with the
repulsive character of $\overline{V}_{\rm 3N}$ in the $^1$S$_0$
channel~\cite{lesinski10a,Hebeler:2009iv}. Although the impact of the
3N contribution is rather insensitive to the structure of the
particular nucleus under consideration, it displays a slight isovector
trend along isotopic (isotonic) chains, most prominently seen in
Fig.~\ref{fig:gaps:coulomb} for the neutron-rich lead isotopes~\cite{lesinski10a}.

The main conclusions of the present work are that (i) it is essential
to include 3N contributions to the pairing interaction for a quantitative
description of nuclear pairing gaps, (ii) the first-order low-momentum
results leave about $30 \%$ room for contribution from higher orders,
e.g., from the coupling to (collective) density, spin and isospin
fluctuations, (iii) in the next steps, the normal self-energy and
higher-order contributions to the pairing kernel must be computed
consistently starting from the same low-momentum NN and 3N
interactions.

\vspace*{1mm}
\noindent
This work was supported in part by the U.S. Department of Energy under
Contract Nos. DE-FG02-96ER40963 and DE-FG02-07ER41529 (University of
Tennessee), the Natural Sciences and Engineering Research Council of
Canada (NSERC), and by the Helmholtz Alliance Program of the Helmholtz
Association, contract HA216/EMMI ``Extremes of Density and
Temperature: Cosmic Matter in the Laboratory''.  TRIUMF receives
federal funding via a contribution agreement through the National
Research Council of Canada.


\begin{thebibliography}{10}
\bibitem{bender03a}
M. Bender, P.-H. Heenen and P.-G. Reinhard, Rev. Mod. Phys. 75 (2003) 121.
\bibitem{duguet08a}
T. Duguet and T. Lesinski, Eur. Phys. J. Special Topics 156 (2008) 207.
\bibitem{lesinski09a}
T. Lesinski, T. Duguet, K. Bennaceur and J. Meyer, Eur. Phys. J. A40
(2009) 121.
\bibitem{duguet09a}
T. Duguet and T. Lesinski, AIP Conf. Proc. 1165 (2009) 243, arXiv:0907.1043.
\bibitem{bogner03a}
S. K. Bogner, T. T. S. Kuo and A. Schwenk, Phys. Rep. 386 (2003) 1.
\bibitem{bogner07a}
S. K. Bogner, R. J. Furnstahl, S. Ramanan and A. Schwenk, Nucl. Phys.
A784 (2007) 79.
\bibitem{lesinski10a}
T. Lesinski, K. Hebeler, T. Duguet and A. Schwenk, in preparation.
\bibitem{bogner09a}
S. K. Bogner, R. J. Furnstahl, A. Nogga and A. Schwenk, arXiv:0903.3366.
\bibitem{Hebeler:2009iv}
K. Hebeler and A. Schwenk,  arXiv:0911.0483.
\bibitem{hebeler09b}
K. Hebeler, T. Duguet, T. Lesinski and A. Schwenk, Phys. Rev. C80
(2009) 044321.
\bibitem{duguet02}
T. Duguet, P. Bonche, P.-H. Heenen and J. Meyer, Phys. Rev. C65 (2001)
014311.
\end{thebibliography}
\end{document}